\documentclass[a4paper, amsfonts, amssymb, amsmath, reprint, showpacs, showkeys, nofootinbib,prl, twoside, floatfix]{revtex4-2}
\usepackage[english]{babel}
\usepackage[utf8]{inputenc}
\usepackage{upgreek}
\usepackage[colorinlistoftodos, color=green!40, prependcaption]{todonotes}
\usepackage{amsthm}
\usepackage{mathtools}
\usepackage{siunitx}
\usepackage{xcolor}
\usepackage{graphicx}
\usepackage[autostyle]{csquotes}
\usepackage[left=16mm,right=16mm,top=35mm,bottom=20mm,columnsep=14pt]{geometry} 
\usepackage{adjustbox}
\usepackage{placeins}
\usepackage{multirow}
\usepackage[T1]{fontenc}
\usepackage{lipsum}
\usepackage{url}
\usepackage{natbib,hyperref}
\hypersetup{colorlinks=true, urlcolor=blue, linkcolor=blue, citecolor=blue}
\usepackage{multirow, booktabs}
\usepackage{epstopdf}

\begin{document}
\renewcommand{\figurename}{FIG.{}}
\title{Observation of Resonant Tunneling from Molecular Shape into Vibronic Feshbach Resonances Followed by Mode-Specific Fragmentation}

\author{Narayan Kundu\textit{$^{1,2}$}}
\email{kundu.narayan1995@gmail.com}

\author{Meenakshi Rana\textit{$^{3}$}}

\author{Aryya Ghosh\textit{$^{3}$}}
\email{aryya.ghosh@ashoka.edu.in}

\author{Dhananjay Nandi\textit{$^{1,4}$}}%
\email{dhananjay@iiserkol.ac.in}

\affiliation{$^1$Indian Institute of Science Education and Research Kolkata, Mohanpur-741246, India.\\ $^2$University of Kassel, Institute of Physics, Heinrich-Plett-Str. 40, 34132 Kassel, Germany\\ $^3$Department of Chemistry, Ashoka University, Sonipat, Haryana, 131029, India.\\ $^4$Center for Atomic, Molecular and Optical Sciences $and$ Technologies, Joint initiative of IIT Tirupati $and$ IISER Tirupati, Yerpedu, 517619, Andhra Pradesh, India}




\begin{abstract}
We present a kinematically complete study of dissociative electron attachment (DEA) in linear OCS molecules, focusing on how electrons resonantly attach and trigger dissociation. Near the Franck-Condon regime, DEA is dominated by molecular shape resonances, where transient OCS$^-$ states form with high vibrational amplitudes, spectroscopically evident as broad features in DEA cross-sections. As the electron beam energy increases from 5.5 to 6.0 eV, S$^-$ population shifts from lower to higher-energy highly dense bending vibrational states, reinforcing our findings on dipole-forbidden vibronic intensity borrowing. Our advanced potential energy curve calculations, employing the Equation-of-motion coupled cluster singles and doubles for electron attachment (EA-EOMCCSD) method, reveal that beyond the shape resonance, non-adiabatic resonant tunneling governs the avoided crossings, dynamically generating three mode-specific vibronic Feshbach resonances before complete dissociation into three distinct kinetic energy bands of S$^-$. Our theoretical results probe most of the experimental observations quantitatively and qualitatively. These insights deepen our fundamental understanding of resonance-mediated dissociation in electron-molecule resonant scattering, with broader implications for quantum mechanics, plasma physics, vibrational revival, astrochemistry, and radiation damage researches. 
\end{abstract}

\maketitle
The interaction of low-energy electrons (< 50 eV) with molecules plays a crucial role in plasma physics \cite{lee_plasma_prl,demidov2015measurements}, prebiotic astrochemistry \cite{sandford2020prebiotic,arumainayagam2019extraterrestrial}, surface sciences \cite{arumainayagam2010low}, and radiation-induced damage in biological systems \cite{boudaiffa2000resonant,pan_dna_prl}. Dissociative Electron Attachment (DEA) is a key mechanism where an incident low energy electron (< 15 eV) transiently binds to a neutral molecule, forming a short-lived anionic state that dissociates into charged and neutral fragments. It is a complex resonant process in which a strong interaction exists between electronic and nuclear motion \cite{krishnakumar2018symmetry,dea_hnco_prl,fabrikant2017recent}, attributed DNA damage \cite{boudaiffa2000resonant,pan_dna_prl} to molecular cluster formation \cite{illenberger_chem_rev}. The complexity increases from diatomic to polyatomic molecules, making the DEA's theoretical description more challenging \cite{fabrikant2017recent}. Using one dimensional non-local model, Martin \v{C}\'{i}\v{z}ek \textit{et al.} reported a few \cite{cizek_pra_co2_22,cizek_prl_hcno,dea_vibronic_cizek1} to probe the vibronic structures of DEA dynamics in electron-molecule resonance continuum. Thus, it is desirable for benchmarking the theory to a few diatomic and polyatomic molecules to understand our rich experimental observations in recent times \cite{kundu2024observation}. 

In this study, we report a kinematically complete investigation of DEA dynamics in linear triatomic carbonyl sulfide (OCS) molecules, elucidating the interplay among fundamental mechanisms of molecular shape resonances \cite{moradmand2013dissociative}, resonant tunneling \cite{resonant_tunneling_prl}, and vibronic Feshbach resonances \cite{vfr_uracil,vfr_kang_prl}, which collectively drive the mode-specific fragmentation \cite{mode_specific_pamir_prl,mode_specific_tarana_pra}. To validate our experimental findings, we performed high-accuracy potential energy curve calculations employing the Equation-of-motion coupled cluster singles and doubles for electron attachment (EA-EOMCCSD) \cite{RevModPhysbartlett} method. Importantly, EA-EOMCCSD is specifically designed to address the complex challenges of DEA dynamics due to  its enormous capability  of including dynamic and non-dynamic electron correlation effect which helps to model electron detachment processes correctly \cite{ghosh2012equation}. Therefore, for DEA studies, particularly when dealing with transient anions and non-adiabatic transitions, EA-EOMCCSD method is the most robust and accurate one \cite{ghosh2013cap}.  

Additionally, we employed the multi-configurational time-dependent Hartree (MCTDH) \cite{physrepmctdh,cplmctdh1990} method combined with excited-state wavefunction analysis, providing a precise framework for probing the dynamics of electronically excited molecular systems. This approach is particularly effective in capturing nonadiabatic couplings and multi-configurational effects, as exemplified by the OCS$^{-}$ transient anion system. By integrating these theoretical advancements with velocity map imaging (VMI) experiments, we achieve a comprehensive, angle-, channel-, and time-resolved understanding of the dissociation dynamics, bridging electronic structure, nuclear motion, and decay pathways \cite{Mikosch_prl_2013}.

\begin{figure*}
    \centering
    \includegraphics[scale=0.82]{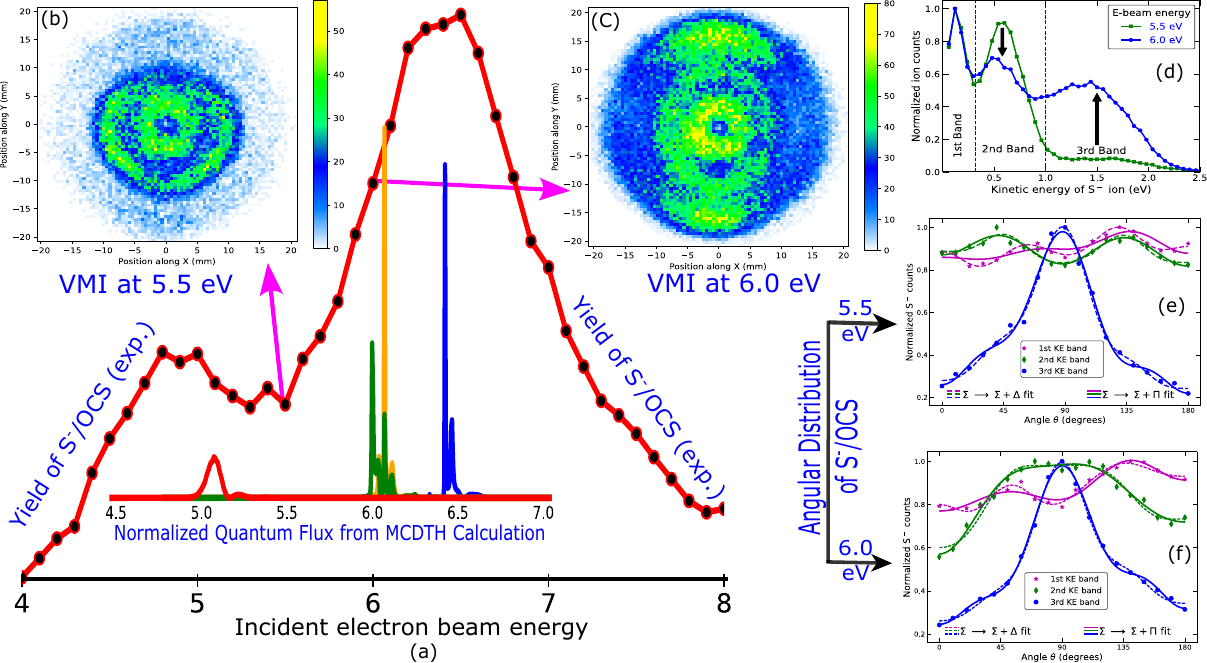}
    \caption{\footnotesize {(a) The normalized quantum flux yield of S$^{-}$/OCS, calculated theoretically, is compared with the experimentally obtained excitation function or ion yield of S$^{-}$/OCS as a function of beam energy. (b, c) Time-gated (50 ns) conical wedge slice images of S$^{-}$ are shown at beam energies of 5.5 eV and 6.0 eV, respectively. (d) The kinetic energy distribution of the S$^{-}$ product demonstrates that an increase in beam energy leads to a higher number of S$^{-}$ products in the third kinetic energy (KE) band. Additionally, the peak positions of the second and third KE bands shift to higher energy values as beam energy increases. (e, f) Angular distributions for each KE band are presented and fitted using the O'Malley and Taylor expression \cite{omelly_1968}. Solid and dashed lines indicate the best fit for $\Sigma \longrightarrow \Sigma + \Pi/\Delta$ symmetric transition states over the experimental data points.}}
    \label{fig:fig1}
\end{figure*}

The experiment utilized a 100 ns pulsed electron source coupled to a home-built velocity map imaging (VMI) spectrometer housed in an ultra-high vacuum (UHV) chamber with a base pressure of \(10^{-9}\)~mbar. Ion detection was achieved using a Z-stack of three microchannel plates (MCPs) paired with a hexanode delay-line detector, enabling precise \textit{x-y} position and time-of-flight (TOF) measurements of each event. During measurements, the UHV pressure was maintained at \(2 \times 10^{-7}\)~mbar to ensure single-collision conditions. Ion events were recorded in list mode format (lmf) using CoboldPC software (RoentDek GmbH) \cite{ullmann1999list}. Offline analysis involved applying a TOF window to the cylindrically symmetric raw data to generate time-gated velocity slice images (VSI), kinetic energy (KE) distributions, and angular distributions (AD), providing complete kinematic information. To mitigate artifacts from the standard parallel slicing method—which uses a TOF window with a 10–25\% velocity spread around the momentum distribution center and exaggerates low-momentum fragments \cite{nag_nccn,slaughter_2013}—we employed the conical time-gated wedge slice imaging technique \cite{kundu2023breakdown,kundu2024observation} for accurate momentum and KE reconstruction. Electron beam energies and fragment KE were calibrated against the well-characterized O$^-$/O$_2$ resonance at 6.5 eV \cite{nandi2005velocity,nandi_ke}.

We began our experiment by recording the TOF-based mass spectra to confirm the masses of the product ions, as reported in our previous study \cite{kundu2024observation}. Subsequently, we selected the relevant TOF mass region to investigate the excitation function of the S$^-$ fragment as a function of the electron beam energy. As illustrated in the ion-yield spectrum in FIG. \ref{fig:fig1}(a), it is evident that the formation of S$^-$ exhibits two distinct resonant features, with peak positions at 5.0 eV and 6.5 eV beam energies. These resonant peaks are consistent with previous DEA studies on gas-phase OCS molecules \cite{iga1995,iga1996_elsever}. A noteworthy observation from the S$^-$/OCS ion yield is that as the electron beam energy shifts from the 5.0 eV to the 6.5 eV resonance points, the formation of S$^-$/OCS increases significantly. In this study, we report a detailed and comprehensive analysis of the dynamics underlying the sudden increase in DEA cross-sections associated with electron attachment to a triatomic molecule. Additionally, the ion-yield of S$^-$/OCS reveals a broader resonance feature near 5.0 eV compared to the narrower resonance observed at 6.5 eV, as rationalized below.  

\begin{figure*} [htb]
    \centering
    \includegraphics[scale=0.85]{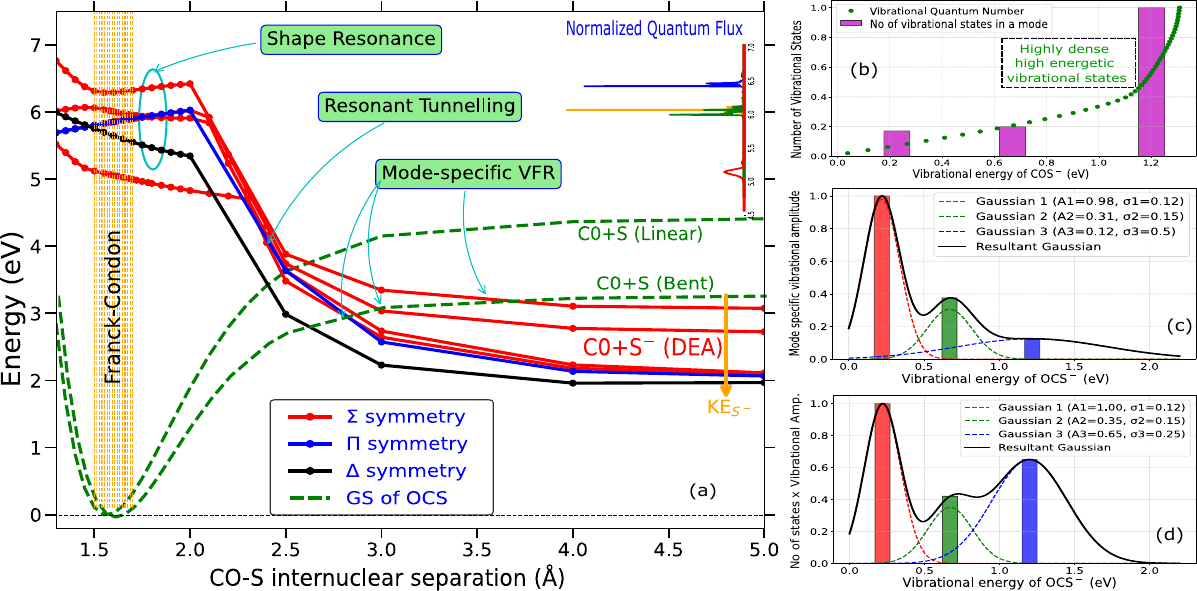}
    \caption{\footnotesize{(a) Calculated potential energy curves for the ground states and select excited states of OCS, where resonant electron attachment occurs in the Franck-Condon (FC) region. The state symmetries are indicated in the legend. The normalized quantum flux, derived from time-resolved Gaussian wave packet analysis, is shown in the top-right corner. (b) Calculated vibrational energy eigenvalues are divided into three distinct vibrational energy bands. The number of vibrational states in the most populated mode is normalized. (b and c) The distribution of vibrational states in a given mode is fitted with three Gaussian functions. In the first case, mode-specific vibrational amplitudes are considered. In the second case, these amplitudes are multiplied by the number of accessible vibrational states in the mode. }}
    \label{fig:fig2}
\end{figure*}

After determining the desired resonance energy, we adjusted the beam energy accordingly to capture the momentum images. FIG. \ref{fig:fig1}(b,c) shows the conical time-gated (50 ns) wedge slice images of S$^-$/OCS at beam energies of 5.5 eV and 6.0 eV. We focused on these two energy levels because they are crucial for understanding the changes in the DEA process. From FIG. \ref{fig:fig1}(b,c), it is clear that when the beam energy increases from 5.5 eV to 6.0 eV, the production of S$^-$ with higher momentum significantly rises in the outermost (third) momentum band. A closer look at the conical slice images shows that this increase in high-momentum S$^-$ production is accompanied by a gradual decrease in S$^-$ fragmentation within the second momentum band. However, the increase in high-momentum S$^-$ production is much greater than the decrease seen in the second momentum band. FIG. \ref{fig:fig1}(d) displays the kinetic energy (KE) distribution of S$^-$, derived from the slice images, offering a clearer view of how S$^-$ formation changes across different momentum bands as the beam energy varies. If we carefully examine the KE distribution as a function of beam energy, we notice that the peak populations of the S$^-$ product in the second and third KE bands shift toward higher energy values.   

We then extracted the angular distribution (AD) for each KE band and fitted the data using the O’Malley and Taylor equation \cite{omelly_1968}. This method allowed us to identify the symmetry of the resonant states near the vertical FC regime, where the electron attaches resonantly to one of the unoccupied molecular orbitals of OCS. FIG. \ref{fig:fig1}(e-f) shows the fitted curves for the KE band-integrated ADs of S$^-$, utilizing the C$_{\infty V}$ point group symmetries. The fitting results indicate that the $\Sigma \longrightarrow \Sigma$ symmetric transitions predominantly describe the angular momentum bands. However, the most accurate representation was achieved when using AD models that include $\Sigma \longrightarrow \Sigma + \Pi$ (solid curves) or $\Sigma \longrightarrow \Sigma + \Delta$ (dashed lines) symmetric transitions, as shown in Fig. \ref{fig:fig1}(e-f). A comprehensive analysis of the angular distribution of S$^-$ fragments has been previously reported \cite{kundu2024observation}. So far, by synthesizing experimental findings—including ion-yield measurements, slice imaging, kinetic energy profiles, and angular distribution analyses—with molecular symmetry principles, we have systematically demonstrated how electron beam energy influences transient resonance behavior. This interaction directly affects the momentum characteristics of the reaction products. Through this integrated approach, we illustrate how precise energy modulation in electron beams acts as a crucial factor in determining the dynamic pathways and final outcomes of molecular interactions.

We performed quantum mechanical calculations to obtain the ground-state potential energy curves (PECs) of OCS using the coupled cluster singles and doubles (CCSD) method. Additionally, we calculated the PECs of several excited electron-attached states of neutral OCS using the EA-EOMCCSD method \cite{eomcc3_open,eomcc1}, as shown in FIG. \ref{fig:fig2}(a). In our calculations, we employed the aug-cc-pVTZ basis set for all atoms. The PECs of the OCS molecule provide significant insights into dissociation dynamics as a function of the CO-S internuclear distance. At short distances (less than 2 \AA), the ground state (GS) of OCS indicates a stable molecular configuration with low potential energy. As the bond length increases, the potential energy curves representing the dissociated fragments (CO + S) become prominent, indicating the weakening of the CO-S bond and the transition to a dissociated state. The plot highlights the presence of resonances, including shape resonances and vibronic Feshbach resonances (VFR). Near the Franck-Condon (FC) region, shape resonances appear, where the incident electron temporarily attaches to the molecule without significantly changing its geometry. This transient resonance stabilization forms a quasi-bound state that may either return to the neutral molecule or proceed to dissociation. As the bond length increases further, non-adiabatic transitions emerge, leading to resonant tunneling. This process involves avoided crossings between PECs and results in the formation of Feshbach resonances, which redistribute vibrational energy before dissociation.

In the FC region, an incident electron can resonantly attach to an unoccupied molecular orbital. At short CO-S internuclear separations (less than 2 \AA), these electron attachments are primarily driven by molecular shape resonances, leading to the formation of transient OCS$^{-}$ with large vibrational amplitudes. Near the FC region, FIG. \ref{fig:fig1}(a) shows a minimum at 6.5 eV in the $^{2}\Sigma$ symmetric excited state of anionic OCS, indicating the presence of stable vibrational states within the open-shell configurations. Our calculations reveal that the incident electron resonantly attaches to the lowest unoccupied molecular orbital (LUMO) for the $^{2}\Delta$ symmetric states of OCS, to LUMO+3 for the $\Pi$ symmetric states, and to LUMO+4,5,6,7 for the four $^{2}\Sigma$ symmetric states of OCS, as indicated by the black, blue and red curves, respectively, in FIG. \ref{fig:fig2}(a). Detailed visualizations and discussions of these molecular orbitals (MOs) can be found in the supporting documents. These molecular shape resonances, caused by resonantly attached electrons with adequate resonance widths, eventually decay into broad vibronic states. This process results in broad DEA cross-section features in the experimental ion yield of S$^{-}$/OCS.

Beyond the FC region, non-adiabatic resonant tunneling governs the transitions of the transient anions, dynamically forming three distinct vibronic mode-specific Feshbach resonances. These transitions occur through resonant tunneling via avoided crossings, leading to the three characteristic kinetic energy bands of S$^{-}$ fragments superimposed over the predissociation continuum. This results rationales with our experimental observation from the time-gated velocity slice images shown in FIG. \ref{fig:fig1}(b,c). Before the transient OCS$^{-}$ fully dissociates at a long range, these resonances redistribute its vibrational energy among three distinct modes, as discussed below.

Moreover, a key observation is the change in yield when the beam energy increases from 5.5 to 6.0 eV. This increase enhances electron attachment due to overlapping resonances near 6.0 eV, resulting in a higher population of high-energy S$^{-}$ fragments. The increase in beam energy activates higher vibrational states, leading to more efficient dissociation. Symmetry considerations also play a role, as the dissociation path changes from linear to bent with increasing CO-S distance. This symmetry transition affects the fragmentation process, influencing the kinetic energy distribution of the S$^{-}$ products. Overall, the PECs illustrate how electron energy and molecular geometry jointly dictate the dissociation pathways and product distributions.

We calculated the vibrational energy eigenfunctions of the transient OCS$^{-}$ anion using the \textit{ab-initio} method and found that the system exhibits anharmonicity. A brief discussion on anharmonicity and vibrational energy eigenvalues is provided in the supplementary section. We noticed cubic fit of vibrational quantum number (\( E_n = E_0 + \alpha n + \beta n^2 + \gamma n^3 \)) in the energy expansion significantly improves fit accuracy over the quadratic fit (\( E_n = E_0 + \alpha n + \beta n^2 \)) and allows estimation of vibrational revival ($\approx$ 8.2 picoseconds) and super-revival times of  the transient OCS$^{-}$ anion \cite{bluhm_revival,vraking_revival}.  Consequently, the vibrational energy density becomes significantly high for the large-amplitude vibrations of OCS$^{-}$, as depicted in FIG. \ref{fig:fig2}(b). At low beam energy (5.5 eV), the high-energy, densely packed vibrational states of the transient anion remain mostly inactive. However, a slight increase in beam energy from 5.5 to 6.0 eV raises the center-of-mass energy of the transient negative ion state (OCS$^{-}$), thereby activating these high-energy vibrational states. This activation can occur through two primary mechanisms: (1) borrowing vibrational intensity \cite{orlandi1973theory} from low-energy vibrational levels, consistent with our observations from the normalized KE distribution of S$^{-}$, and (2) an increase in the number of electron attachments in FC regime. In this specific case, both mechanisms contribute to the enhanced formation of high-energy S$^{-}$ at 6.0 eV beam energy.

This study offers a detailed mechanistic understanding of dissociative electron attachment (DEA) dynamics in carbonyl sulfide (OCS) by combining advanced quantum calculations with complete kinematic experiments. Using velocity map imaging (VMI) and high-level theoretical methods—such as EA-EOMCCSD potential energy curves (PECs) and MCTDH wavepacket dynamics—we explore the roles of molecular shape resonances, non-adiabatic resonant tunneling, and vibronic Feshbach resonances (VFR) in OCS dissociation. Our findings show that electron attachment in the Franck-Condon (FC) region generates transient OCS$^{-}$ via shape resonances, with symmetry-dependent coupling to unoccupied orbitals (LUMO for $^2\Delta$, LUMO+3 for $^2\Pi$, and LUMO+4–7 for $^2\Sigma$ states). These resonances decay into broad vibronic states, leading to the observed DEA cross-sections in S$^{-}$/OCS ion yields. Beyond the FC region, resonant tunneling through avoided crossings drives mode-specific VFR formation, redistributing vibrational energy before dissociation and producing distinct kinetic energy (KE) bands of S$^{-}$ fragments. A key finding is the beam-energy-dependent activation of high-energy S$^{-}$ products. At 5.5 eV, low-energy vibrational states dominate, while increasing the energy to 6.0 eV activates anharmonic, densely packed states via intensity borrowing and enhanced electron attachment. This shift corresponds to the abrupt increase in S$^{-}$ population in the third KE band, linked to overlapping resonances near 6.0 eV. Angular distribution analysis confirms symmetry-driven transitions ($\Sigma \rightarrow \Sigma + \Pi / \Delta$), consistent with $C_{\infty}V$ point group dynamics.

The combination of theory and experiment reveals how electron energy and molecular geometry jointly determine dissociation pathways. EA-EOMCCSD models transient anion states and non-adiabatic couplings, while MCTDH-CAP quantifies quantum flux, matching experimental ion yields. These insights enhance the understanding of DEA in polyatomic systems, highlighting the significance of resonance overlap and vibrational anharmonicity revivals. This letter connects complex electronic structure with nuclear dynamics couplings, and dissociation outcomes, offering a benchmark for modeling electron-driven processes in fields like quantum mechanics, astrochemistry, plasma physics, and radiation damage. By uncovering the role of resonance interactions and vibrational coupling, we provide a foundation for controlling molecular fragmentation through targeted energy modulation.

\textit{Acknowledgement-} N. K. gratefully acknowledges the financial support from `DST of India' for the "INSPIRE Fellowship" program. We recognize Dr Kousik Samanta, and Karunamoy Rajak for their scientific discussion that guides us to reach the goal. We acknowledge financial support from the Science and Engineering Research Board (SERB) for supporting this research under  Project No. "CRG/2019/000872". M. R. and A. G. acknowledges financial support from SERB (SRG/2022/001115).

\bibliography{nk_bib} 
\end{document}


\renewcommand{\figurename}{FIG. S{}}

\title{Observation of Resonant Tunneling from Molecular Shape into Vibronic Feshbach Resonances Followed by Mode-Specific Fragmentation}

\author{Narayan Kundu\textit{$^{1,2}$}}
\email{kundu.narayan1995@gmail.com}

\author{Meenakshi Rana\textit{$^{3}$}}

\author{Aryya Ghosh\textit{$^{3}$}}
\email{aryya.ghosh@ashoka.edu.in}

\author{Dhananjay Nandi\textit{$^{1,4}$}}%
\email{dhananjay@iiserkol.ac.in}

\affiliation{$^1$Indian Institute of Science Education and Research Kolkata, Mohanpur-741246, India.\\ $^2$University of Kassel, Institute of Physics, Heinrich-Plett-Str. 40, 34132 Kassel, Germany\\ $^3$Department of Chemistry, Ashoka University, Sonipat, Haryana, 131029, India.\\ $^4$Center for Atomic, Molecular and Optical Sciences $and$ Technologies, Joint initiative of IIT Tirupati $and$ IISER Tirupati, Yerpedu, 517619, Andhra Pradesh, India}



\maketitle

This supplementary material provides additional information on various aspects of the main manuscript. In summary, these extra details mainly cover the following points.\\

\textbf{(I)} Angular distribution modeling. \\
\  \  \textbf{(II)} The potential energy curves of OCS using two distinct point group Symmetry \\
\  \  \textbf{(III)} Lowest Unoccupied Molecular orbitals of OCS where incident electron resonantly attach\\
 \ \  \textbf{(IV)} Vibrational Energy Analysis of transient OCS Anion in Excited Bent-State and Vibrational revival time.\\
 \ \  \textbf{(V)} Population dynamics and avoided crossing between excited states.
\clearpage

\section{Angular Distribution Modeling}

We have extracted the kinetic energy (KE) band-integrated angular distributions of S$^-$ products. The angular intensity of S$^-$ is calculated by referencing the electron beam direction as the 0$^\circ$ angle, and then normalized to the 90$^\circ$ angle. Given that the dissociation channel produces two fragments, S$^-$ and CO, with nearly equal masses, it is feasible to investigate the fundamental dynamics as a diatomic resembling a two-particle molecular fragmentation process. This analysis considers the symmetries of the C$_{\infty V}$ point group and assumes that the axial recoil approximation is a valid assumption. It is insightful to note that the axial recoil approximation holds fine when diatomic molecular ions are formed in a dissociative state, but falls short if the metastable TNI possesses highly excited bending vibrations.

The angular distribution of S$^-$ /OCS fragments was fitted using the formula from O'Malley and Taylor's work \cite{omelly_1968}. In our previous report \cite{kundu2024observation}, we employed the identical model to analyze the angular distributions of S$^-$ /OCS in a linear triatomic system. The formula used is:

\begin{equation}
\rm{I}(\theta) = \sum_{\mid \mu \mid} \left\lvert \sum_{j = \mid \mu \mid}^{\infty} a_{j}\ e^{i\delta_{j}} \ Y_{j, \mu}(\theta, \phi)\right\lvert^{2}
\end{equation}

Here, the function I$(\theta)$ represents the angular dependence, whereas the summing over $\mu$ accounts for the contribution of various states in the given process. The values of $\mu$ = 0, 1, and 2 are associated with the transitions to the $\Sigma$, $\Pi$, and $\Delta$ symmetric final states concerning the ground state. The coefficients $a_{j} $'s are the relative weighting factors of the different partial waves, and $\delta_{j} $'s denote the phase differences of each partial wave concerning the lowest-order partial wave responsible for the particular transition.

\begin{figure}[htb!]
\centering
\includegraphics[width=\textwidth]{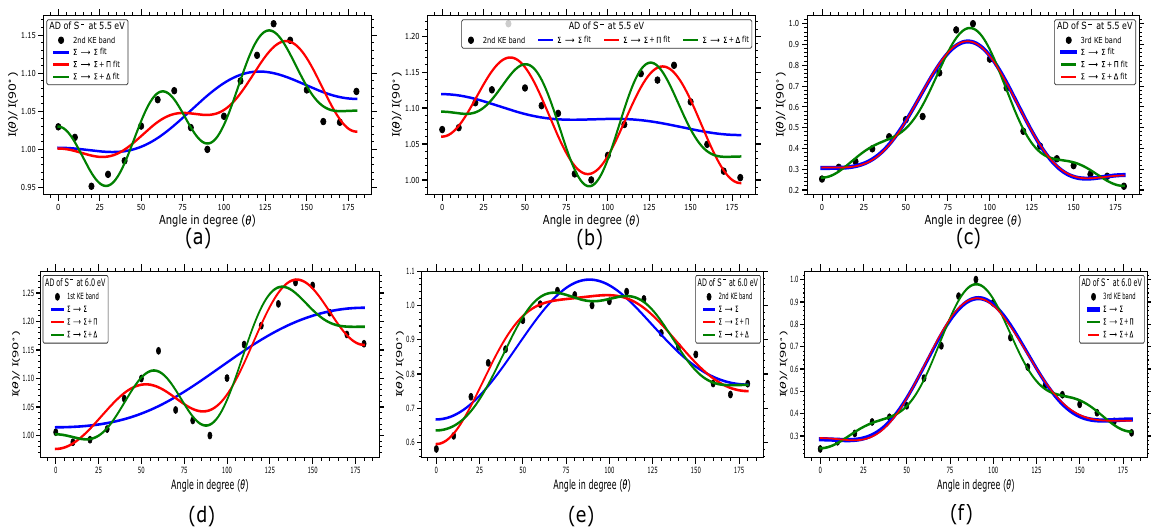}
\caption{Modeling angular distribution of S$^-$ /OCS for its three distinct KE bands using C$_{\infty V}$ point group symmetries for 5.5 eV beam energy (a, b, c) and 5.5 eV beam energy (d, e, f), respectively}
\label{fig:fit_c2v}
\end{figure}

We have modeled the angular distribution of S$^-$ /OCS for its three distinct KE bands using C$_{\infty V}$ point group symmetries for 5.5 eV beam energy (a, b, c) and 5.5 eV beam energy (d, e, f), respectively. The results are shown in FIG. S \ref{fig:fit_c2v}. It is evident from the plots that the $\Sigma \longrightarrow \Sigma$ symmetric transition is a suitable match for all three ADs of S$^-$, but the most accurate fit is achieved when using ADs models with $\Sigma \longrightarrow \Sigma + \Pi$ or $\Sigma \longrightarrow \Sigma + \Delta$ symmetric transitions. The findings imply that the KE bands near the 5.5 and 6.0 eV resonances for S$^-$ /OCS are mostly linked to the $\Sigma$ symmetry association in the OCS$^{-*}$ formation process.

We have also fit the angular distributions of S$^-$ inside the C$_S$ point group symmetries (A$^{\prime}$ and A$^{\prime\prime}$), using the expressions employed by Szymanska et al. \cite{ch3cho_dea_cs}. The anisotropy observed during the A$^{\prime}$ to A$^{\prime}$ transition may be characterized by a linear combination of different powers of cosine. On the other hand, during the A$^{\prime}$ to A$^{\prime\prime}$ transition, the anisotropy can be described by a linear combination of sine functions. The molecule's significantly low level of symmetry is the primary reason behind this. The fit functions are constructed using sine and cosine functions up to the seventh-order exponent.

\begin{figure}[htb!]
\centering
\includegraphics[width=\textwidth]{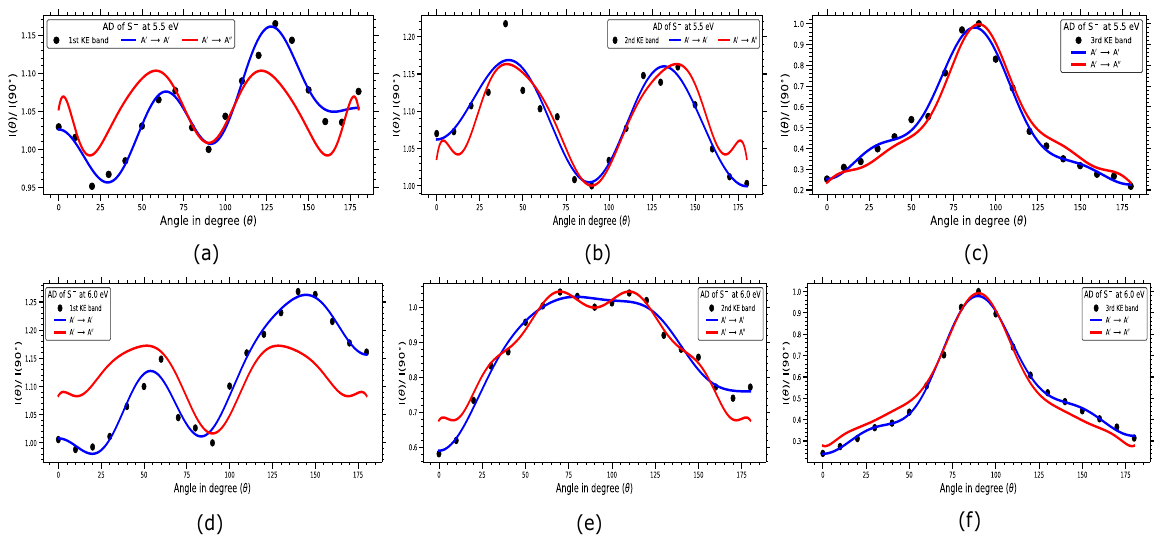}
\caption{Modeling angular distribution of S$^-$ /OCS for its three distinct KE bands using C$_S$ point group symmetries for 5.5 eV beam energy (a, b, c) and 5.5 eV beam energy (d, e, f), respectively}
\label{fig:fit_cs}
\end{figure}

The results of the fit are shown in FIG. S \ref{fig:fit_cs}. It is evident from the plots that the A$^{\prime}$ $\longrightarrow$ A$^{\prime}$ transition fits better than the A$^{\prime}$ $\longrightarrow$ A$^{\prime\prime}$. The fit parameters are provided in the supplementary information. In the highest momentum band of S$^-$, the contribution from A$^{\prime}$ and A$^{\prime\prime}$ symmetries is well behaved. Thus, the DEA dynamics of OCS exhibit a partial correlation between the rotational point group symmetries C$_{\infty V}$ and C$_{s}$. This partial correlation may arise from the linear to bending instability of the temporary OCS$^-$ formation, which can drive the Renner-Teller vibronic splittings of $\Pi$ and $\Delta$ symmetric states.

\section{The potential energy curves of OCS using two distinct point group Symmetry}
\begin{figure}[htb!]
    \centering
    \includegraphics[width=\textwidth]{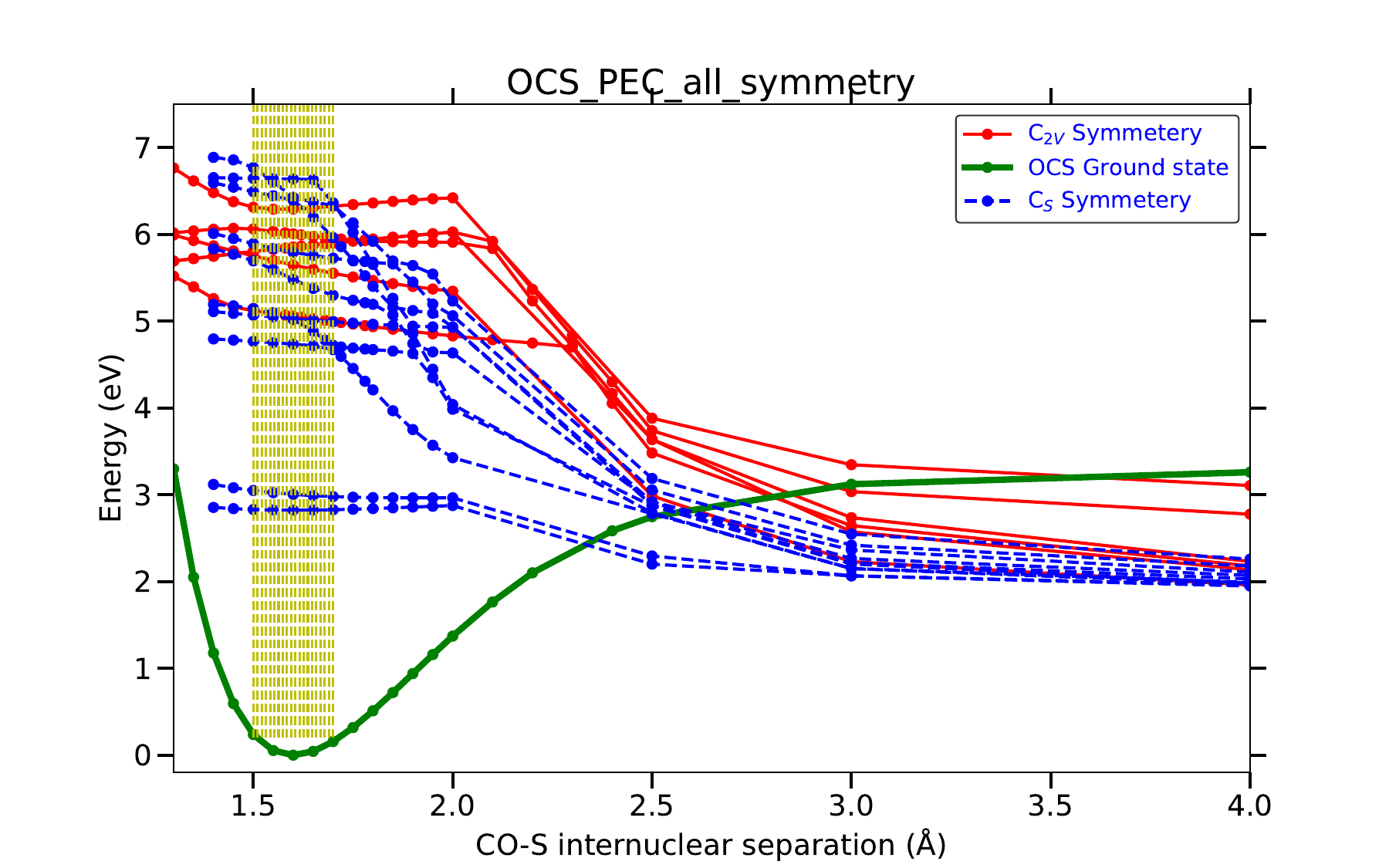}
    \caption{Potential energy curves of OCS calculated under C$_2$v and C$_S$  symmetry frameworks plotted as a function of CO-S internuclear separation.}
    \label{fig:pecs_all}
\end{figure}

The potential energy curves (PECs) of OCS (carbonyl sulfide) calculated using the EA-EOMCCSD method provide significant insights into the effects of symmetry frameworks, specifically C$_2$v and C$_S$, on excited states and fragmentation dynamics. FIG. S \ref{fig:pecs_all} reveals that the C$_S$  symmetry calculations reveal a denser spectrum of dissociative states near the high-momentum bands of S$^-$ / OCS fragmentation, highlighting the nuanced electronic structure captured in this less restrictive framework. A crucial observation is that the OCS bending mode contributes substantially to the formation of high-energy S$^-$, while its contribution is comparatively smaller for the formation of low-momentum S$^-$. These findings underscore the interplay between vibrational modes and fragmentation pathways, with C$_S$ symmetry offering critical details on dissociation dynamics that are not as evident under C$_2$v symmetry constraints.

\section{Lowest Unoccupied Molecular orbitals of OCS where incident electron resonantly attach}
\begin{figure}[!ht]
    \centering
    \includegraphics[scale=0.8]{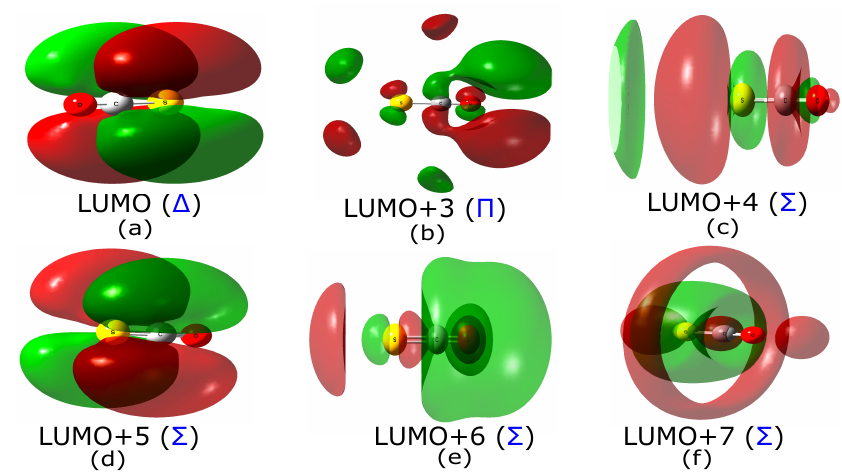}
    \caption{Plots (a,b) reflect the different LUMO of OCS, where incidents electrons are attached. The molecular state symmetry of each incident electron resonantly attached LUMO has been pointed out within the first bracket using blue color.}
    \label{fig:homo_lumo}
\end{figure}

As we discussed, the CCSD level of theory is used to probe our observations. During the calculation of resonance, we used the aug-cc-pVtZ basis set. FIG. S \ref{fig:homo_lumo} reflects the LUMOs of OCS, where incident electrons are resonantly attached.

\section{Vibrational Energy Analysis of transient OCS Anion in Excited Bent-State and Vibrational revival time}
The carbonyl sulfide transient anion (OCS$^-$) is a triatomic molecular species adopting a bent geometry in its electronically excited states. Vibrational energy levels of such excited anions typically exhibit both harmonic and anharmonic characteristics, necessitating precise quantification for elucidating molecular dynamics, spectroscopic signatures, and quantum phenomena. In this study, the vibrational eigenenergies of the OCS$^-$ temporary negative ion were computed using the \textit{ab initio} Møller–Plesset second-order perturbation theory (MP2) to account for electron correlation effects. These eigenvalues, presented in Tab. \ref{tab:tab_1} alongside derived parameters, form the basis for analyzing anharmonicity and wavepacket revival dynamics. For systems with strong anharmonicity, such as the bent excited state of OCS anions, higher-order polynomial fitting is necessary to capture energy level spacing accurately. By applying quadratic and cubic polynomial fits to the vibrational spectrum, we extract revival and superrevival times—key metrics for probing ultrafast nuclear motion in triatomic systems. Vibrational wavepackets in polyatomic systems undergo periodic revival and fractional revival phenomena due to the quantized energy level structure. The study of these revivals provides insight into coherent molecular dynamics and helps characterize potential energy surfaces. This work provides a comprehensive vibrational analysis of the bent OCS$^-$ excited state, bridging \textit{ab initio} quantum chemistry with time-resolved dynamical phenomena.

\begin{longtable}{c c c c c}

\caption{Vibrational energy eigenvalues of the OCS$^-$ anion calculated from MP2 calculations. The table lists the eigenvalues in electronvolts (eV), their corresponding intensities, eigenerrors, and excitations in wavenumbers (cm$^{-1}$) for the first 48 vibrational states. These results provide insight into the vibrational structure and energy level spacings for the triatomic OCS$^-$ anion in its excited states.}\\

\textbf{No.} & \textbf{Eigenvalue} & \textbf{Intensity} & \textbf{Eigenerror} & \textbf{Excitation} \\
            & (eV)                &                     & (eV)                & (cm$^{-1}$)          \\
\midrule
\endfirsthead

\toprule
\textbf{No.} & \textbf{Eigenvalue} & \textbf{Intensity} & \textbf{Eigenerror} & \textbf{Excitation} \\
            & (eV)                &                     & (eV)                & (cm$^{-1}$)          \\
\midrule
\endhead

\midrule
\endfoot

\bottomrule
\endlastfoot

1  & 0.03917  & 3.347E-1 & 3.81E-11 & 315.92  \\
2  & 0.11650  & 2.041E-3 & 3.81E-11 & 939.66  \\
3  & 0.19232  & 1.536E-1 & 3.81E-11 & 1551.14 \\
4  & 0.26656  & 2.318E-3 & 3.81E-11 & 2149.94 \\
5  & 0.33918  & 1.044E-1 & 3.81E-11 & 2735.65 \\
6  & 0.41007  & 2.118E-3 & 3.81E-11 & 3307.44 \\
7  & 0.47909  & 7.765E-2 & 3.81E-11 & 3864.14 \\
8  & 0.54630  & 1.731E-3 & 3.81E-11 & 4406.23 \\
9  & 0.61142  & 5.949E-2 & 3.81E-11 & 4931.42 \\
10 & 0.67422  & 1.234E-3 & 3.81E-11 & 5437.95 \\
11 & 0.73503  & 4.531E-2 & 3.81E-11 & 5928.41 \\
12 & 0.79366  & 6.864E-4 & 3.81E-11 & 6401.26 \\
13 & 0.84953  & 3.489E-2 & 3.81E-11 & 6851.95 \\
14 & 0.90180  & 4.608E-4 & 3.81E-11 & 7273.47 \\
15 & 0.95031  & 2.543E-2 & 3.81E-11 & 7664.76 \\
16 & 0.99516  & 6.039E-5 & 3.81E-11 & 8026.54 \\
17 & 1.03615  & 1.598E-2 & 3.81E-11 & 8357.08 \\
18 & 1.07255  & 1.832E-5 & 3.81E-11 & 8650.72 \\
19 & 1.10305  & 8.391E-3 & 3.81E-11 & 8896.73 \\
20 & 1.12385  & 2.087E-4 & 3.81E-11 & 9064.45 \\
21 & 1.13921  & 1.647E-3 & 3.81E-11 & 9188.39 \\
22 & 1.15261  & 1.012E-3 & 3.81E-11 & 9296.44 \\
23 & 1.16441  & 1.024E-3 & 3.81E-11 & 9391.56 \\
24 & 1.17470  & 9.213E-4 & 3.81E-11 & 9474.61 \\
25 & 1.18422  & 8.614E-4 & 3.81E-11 & 9551.38 \\
26 & 1.19319  & 7.759E-4 & 3.81E-11 & 9623.75 \\
27 & 1.20175  & 7.104E-4 & 3.81E-11 & 9692.75 \\
28 & 1.20991  & 6.671E-4 & 3.81E-11 & 9758.61 \\
29 & 1.21775  & 6.410E-4 & 3.81E-11 & 9821.79 \\
30 & 1.22528  & 6.222E-4 & 3.81E-11 & 9882.59 \\
31 & 1.23257  & 6.022E-4 & 3.81E-11 & 9941.32 \\
32 & 1.23963  & 5.808E-4 & 3.81E-11 & 9998.30 \\
33 & 1.24649  & 5.599E-4 & 3.81E-11 & 10053.63 \\
34 & 1.25315  & 5.399E-4 & 3.81E-11 & 10107.32 \\
35 & 1.25959  & 5.205E-4 & 3.81E-11 & 10159.27 \\
36 & 1.26580  & 5.011E-4 & 3.81E-11 & 10209.37 \\
37 & 1.27176  & 4.797E-4 & 3.81E-11 & 10257.48 \\
38 & 1.27745  & 4.544E-4 & 3.81E-11 & 10303.36 \\
39 & 1.28282  & 4.232E-4 & 3.81E-11 & 10346.65 \\
40 & 1.28779  & 3.827E-4 & 3.81E-11 & 10386.70 \\
41 & 1.29220  & 3.288E-4 & 3.81E-11 & 10422.32 \\
42 & 1.29595  & 2.753E-4 & 3.81E-11 & 10452.52 \\
43 & 1.29914  & 2.348E-4 & 3.81E-11 & 10478.29 \\
44 & 1.30180  & 1.898E-4 & 3.81E-11 & 10499.75 \\
45 & 1.30399  & 1.585E-4 & 3.81E-11 & 10517.37 \\
46 & 1.30588  & 1.446E-4 & 3.81E-11 & 10532.67 \\
47 & 1.30764  & 1.328E-4 & 3.81E-11 & 10546.84 \\
48 & 1.30925  & 1.214E-4 & 3.81E-11 & 10559.85 \\

\label{tab:tab_1}
\end{longtable}

Here, the vibrational energy levels for the OCS anion were obtained from quantum chemical simulations and expressed in electronvolts (eV), then converted to wavenumbers (cm$^{-1}$) using:

\[
E_{\text{cm}^{-1}} = E_{\text{eV}} \times 8065.54
\]

The quantum number \( n \) indexes each vibrational level. Polynomial fits were performed using:

\begin{itemize}
    \item \textbf{Quadratic Fit:} \( E_n = E_0 + \alpha n + \beta n^2 \)
    \item \textbf{Cubic Fit:} \( E_n = E_0 + \alpha n + \beta n^2 + \gamma n^3 \)
\end{itemize}

The coefficients \( \beta \) and \( \gamma \) capture the anharmonic contributions. The revival time \( T_{\text{rev}} \) and superrevival time \( T_{\text{super}} \) were calculated as:

\[
T_{\text{rev}} = \frac{4\pi \hbar}{|\beta|}, \quad T_{\text{super}} = \frac{4\pi \hbar}{|\gamma|}
\]

Here, \( \hbar = 5.3088372 \times 10^3 \, \text{cm}^{-1} \cdot \text{fs} \) is Planck's constant in appropriate units.

The use of a \textbf{cubic polynomial fit} for analyzing vibrational energy levels in bent triatomic molecules such as the OCS$^-$ anion in an excited state is physically well-justified. While the Morse potential provides a more realistic model than the harmonic oscillator by incorporating anharmonicity through a quadratic term, it primarily applies to diatomic or weakly coupled systems. In \textbf{bent triatomic molecules}, especially in excited states, \textbf{vibrational mode coupling}, \textbf{electronic resonances}, and \textbf{nonlinear anharmonic effects} become significant. These effects distort the potential energy surface and introduce deviations that cannot be fully captured by a simple quadratic dependence on vibrational quantum number.

A \textbf{cubic fit}:
\begin{itemize}
    \item Accurately reflects these higher-order anharmonic effects,
    \item Improves residual behavior compared to a quadratic fit,
    \item Provides better modeling of energy level spacing and vibrational revival times.
\end{itemize}

Hence, cubic fitting is both mathematically superior and physically relevant for analyzing vibrational spectra in complex, anharmonic molecular systems like \textbf{OCS$^-$} \cite{bluhm_revival,vraking_revival}.

\begin{figure}
    \centering
    \includegraphics[width=0.95\linewidth]{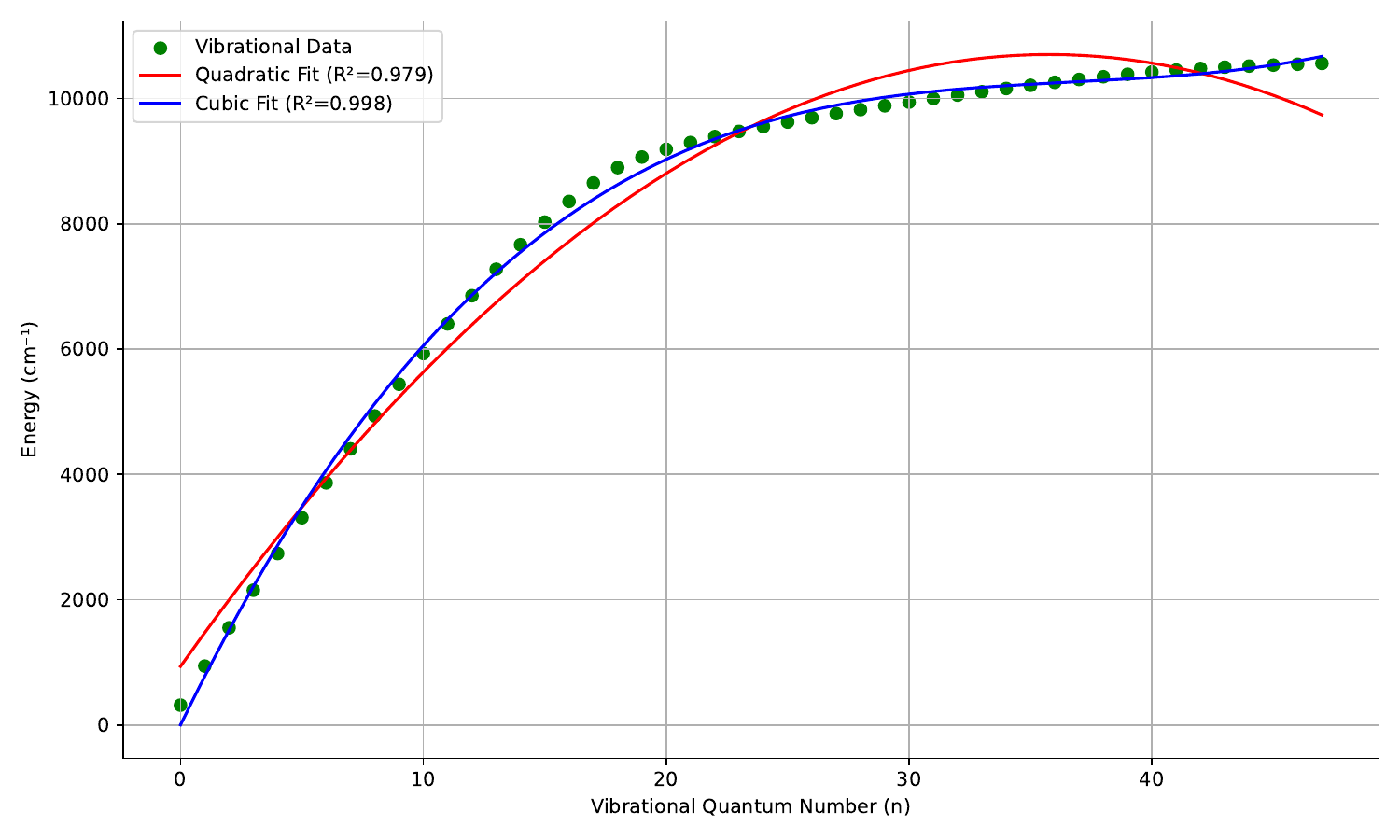}
    \caption{Plot shows the vibrational energy eigenvalues alongside the quadratic and cubic polynomial fits}
    \label{fig:fig1_si}
\end{figure}

\begin{figure}
    \centering
    \includegraphics[width=0.8\linewidth]{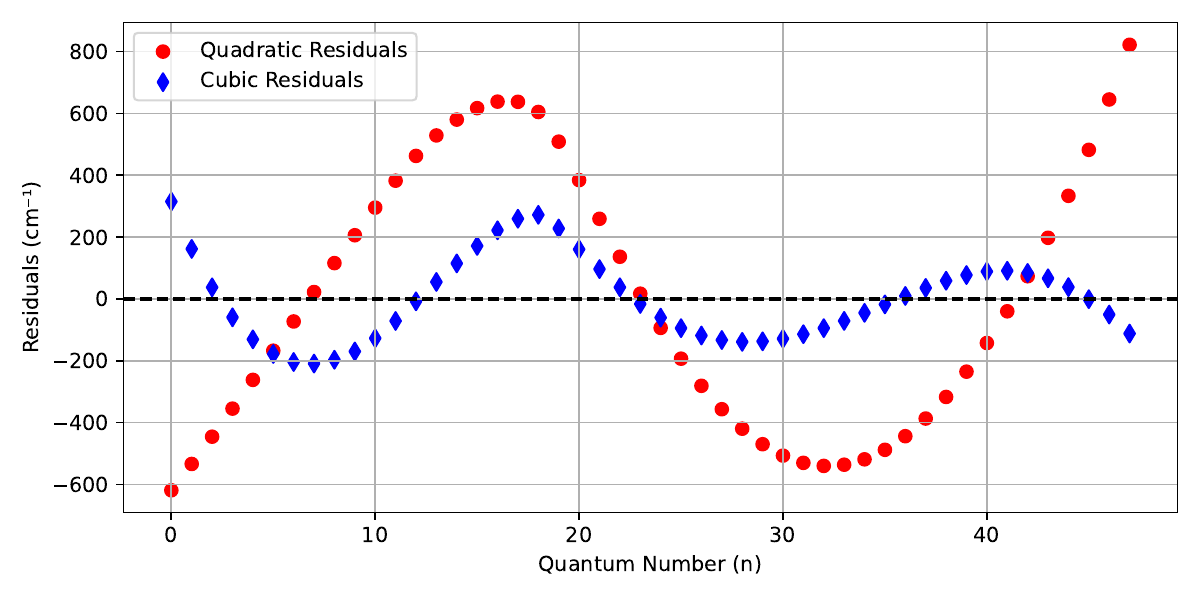}
    \caption{Plot shows the residuals of the fits, showing better performance of the cubic model.}
    \label{fig:fig2_si}
\end{figure}

FIG. S \ref{fig:fig1_si} shows the vibrational energy eigenvalues alongside the quadratic and cubic polynomial fits. The cubic fit provides a substantially better representation of the energy level distribution with an \( R^2 \) of 0.9979, compared to 0.9794 for the quadratic fit. This improvement is further confirmed in the residual plot (FIG. S \ref{fig:fig1_si}), where the cubic fit exhibits minimal deviation across all quantum levels.

The computed timescales are:
\begin{itemize}
    \item \textbf{Revival time:} \( T_{\text{rev}} = 8736.84 \, \text{fs} \)
    \item \textbf{Superrevival time:} \( T_{\text{super}} = 347467.24 \, \text{fs} \)
    \item \textbf{Fractional revivals:} 1/2 at 4368.42 fs, 1/3 at 2912.28 fs
\end{itemize}

\begin{figure}
    \centering
    \includegraphics[width=0.8\linewidth]{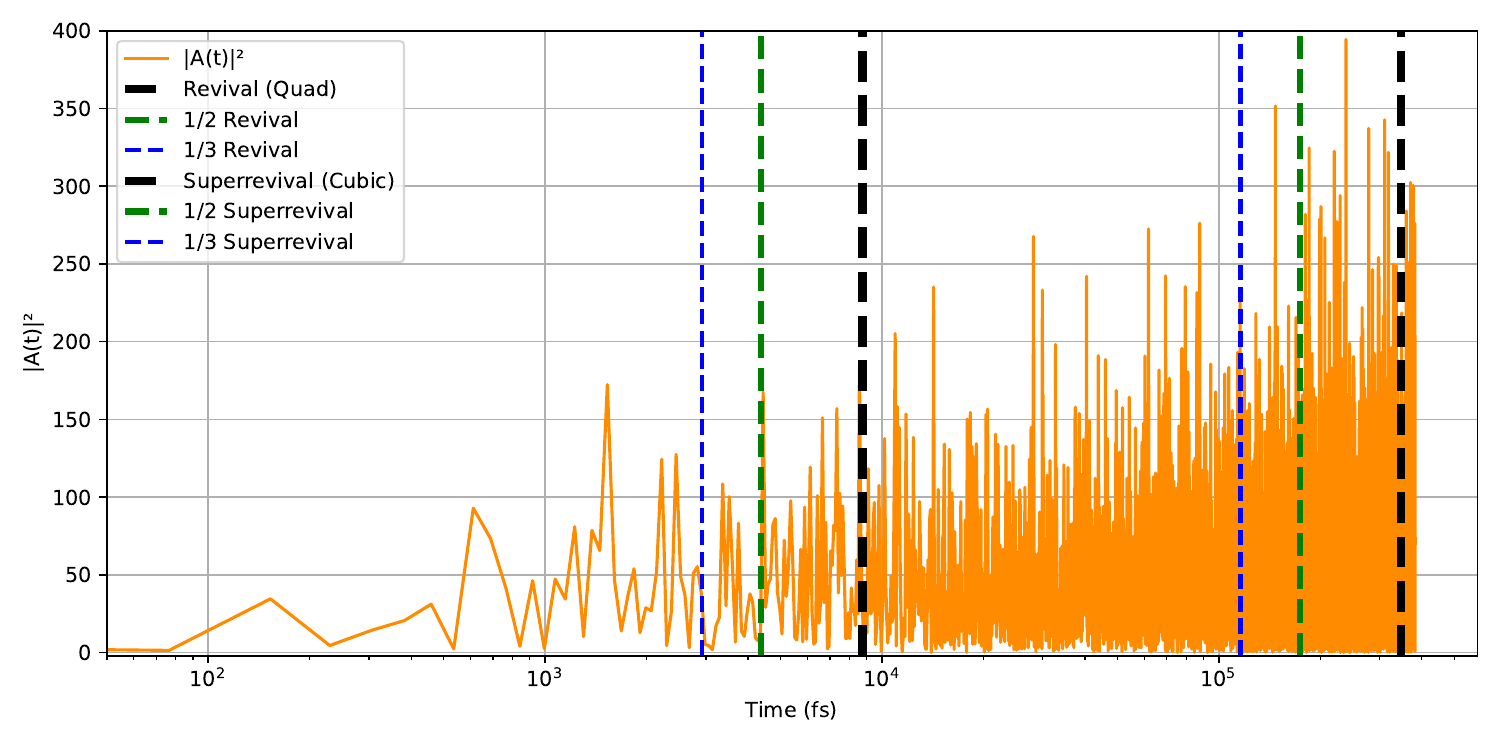}
    \caption{Plot represents the Autocorrelation function \( |A(t)|^2 \) illustrating revival behavior}
    \label{fig:fig3_si}
\end{figure}

\begin{figure}
    \centering
    \includegraphics[width=0.88\linewidth]{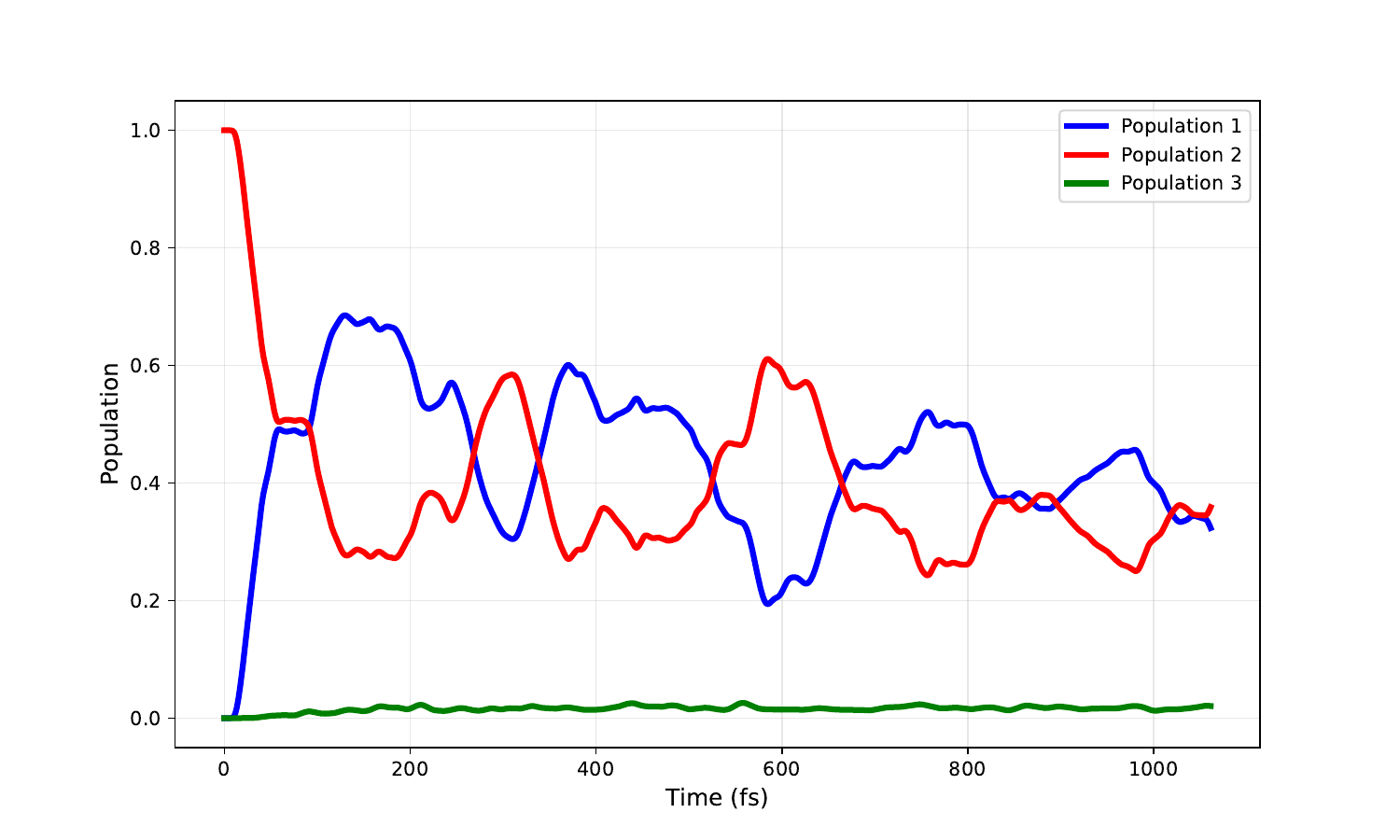}
    \caption{Plot represents population of transient OCS anions as a function of dissociation time scale}
    \label{fig:popula}
\end{figure}

These results illustrate the strong anharmonicity of the excited vibrational potential in OCS anions. FIG. S \ref{fig:fig3_si} presents the autocorrelation function \( |A(t)|^2 \) calculated using the all the vibrational levels. Peaks in the autocorrelation correspond to coherent revivals of the vibrational wavepacket, validating the extracted timescales. This analysis highlights the importance of anharmonic corrections in modeling the vibrational dynamics of bent-state triatomic anions like OCS$^-$. The cubic term in the energy expansion significantly improves fit accuracy and allows estimation of superrevival phenomena. The autocorrelation dynamics reinforce the predicted revival structures and provide a basis for interpreting time-resolved spectroscopic experiments in the ultrafast regime.

\section{Population dynamics and avoided crossing between excited states}
The FIG. S \ref{fig:popula} shows how the populations of different states (population 1, population 2, population 3) evolve over time, measured in femtoseconds (fs). This is a common way to study molecular transitions, especially in ultrafast processes. In the context of OCS, such dynamics could involve vibrational or rotational transitions. OCS molecules exhibit non-statistical vibrational energy relaxation, where energy transitions between modes are uneven and influenced by phase-space structures to revival its vibrational amplitude. 

\bibliography{nk_bib}